# Detection of Hybrid Optical Anapoles in Dielectric Microspheres

*Uttam Manna,\* Iker Gómez-Viloria, Robert Sevik, Isaac Tribaldo, Mahua Biswas, Gabriel Molina-Terriza,\* and Jorge Olmos-Trigo\**

**Nonradiating optical anapoles are special configurations of charge-current distributions that do not radiate. It was theoretically predicted that, for microspheres, electric and magnetic dipolar coefficients can simultaneously vanish by engineering the incident light, leading to the excitation of nonradiating *hybrid* optical anapoles. In this work, the experimental detection of hybrid optical anapoles in dielectric microspheres ($TiO_2$) is reported using dual detection optical spectroscopy, developed to enable sequential measurement of forward and backward scattering under tightly-focused Gaussian beam (TFGB) illumination. The results show that the excitation of $TiO_2$ microspheres (diameter, $d \approx 1$ μm) under TFGB illumination leads to the appearance of scattering minima in both the forward and backward directions within specific wavelength ranges. These scattering minima are found to be due to vanishing electric and magnetic dipolar coefficients associated with hybrid optical anapoles. The ability to confine electromagnetic fields associated with hybrid optical anapoles can give rise to several novel optical phenomena and applications.**

U. Manna, R. Sevik, M. Biswas
Department of Physics
Illinois State University
Normal, Illinois 61709, USA
E-mail: umanna@ilstu.edu

I. Gómez-Viloria, I. Tribaldo, G. Molina-Terriza
Centro de Física de Materiales (CFM-MPC)
CSIC-UPV/EHU
Paseo Manuel de Lardizabal 5, Donostia, San Sebastián 20018, Spain
E-mail: gabriel.molina@ehu.eus

G. Molina-Terriza
IKERBASQUE
Basque Foundation for Science
Bilbao 48013, Spain

J. Olmos-Trigo
Departamento de Física
Universidad de La Laguna
Apdo. 456, San Cristóbal de La Laguna, Santa Cruz de Tenerife E-38200, Spain
E-mail: jolmostr@ull.es

J. Olmos-Trigo
Faculty of Optics and Optometry
Universidad Complutense de Madrid
Madrid 28037, Spain

## 1. Introduction

Optical anapoles are special configurations of charge-current distributions that can confine electromagnetic energy within the particle without radiating in the far-field,[1,2] offering promising applications in nanophotonics.[3–7] It has been predicted that the ability to confine electromagnetic fields[3–6,8–10] can lead to several novel optical phenomena and applications that include enhanced nonlinear effects,[11–15] nanolasers,[16] ideal magnetic scattering,[17] broadband absorption,[18] sensing,[19] metamaterials and metasurfaces,[20] extremely high Q-factors ($\approx 10^6$),[21–24] four-wave mixing,[25] Raman scattering,[26,27] and photothermal nonlinearities.[28]

For time-varying oscillating charge-current distributions, for subwavelength particles, the electrodynamic anapoles are typically related to the superposition of the scattered fields of conventional and toroidal dipoles, given by, $\mathbf{P} = -ik\mathbf{T}$.[8,29] This means that if the contributions of the Cartesian electric ($\mathbf{P}$) and toroidal dipoles ($\mathbf{T}$) to the scattered field are *equal* and *out-of-phase*, the far-field radiation vanishes. In this regime, the optical anapoles are associated with the excitation of single dipolar modes – hence, so far, they have only been observed in nanosized particles. In contrast, for larger particles, these effects are inaccessible due to the contributions from higher-order multipolar modes under plane wave illumination, and the description in terms of $\mathbf{P}$ and $\mathbf{T}$ fails, as the long-wave approximation is no longer valid.

For mesoscopic dielectric spheres, the resonance behavior becomes more complex due to the possibility of multiple radial modes associated within the particle. For example, according to Mie theory, the electric dipole resonance (governed by the Mie coefficient $a_1$) is determined by the zeros of spherical Bessel functions and their derivatives, which define the allowed standing wave patterns inside the sphere. As the size parameter increases, higher-order radial electric dipolar modes, associated with successive zeros of the spherical Bessel functions and characterized by additional internal field nodes, become accessible. These modes can also satisfy the anapole condition, in which interference between internal field components leads to suppression of the far-field dipole radiation.[30–32]

Recently, it was predicted[30,31] that it is possible to unravel dipolar regimes in a homogenous lossless dielectric microsphere in the short wavelength approximation, where the particle diameter, d, is greater than the wavelength of the incident light, λ, under illumination with a pure dipolar field (PDF). More specifically, it was demonstrated that for large particles, illumination of PDF can excite only electric and magnetic dipolar modes.[30,31] This phenomenon stems from the fact that, for particles placed at the focus, the multipolar modes that are not present in the incident beam cannot excite the corresponding modes of the particle, as has been previously demonstrated by tailoring multipolar Mie scattering using Cylindrical Vector Beams[29,33,34] and changing the helicity and angular momentum of the incident beam.[35] Nonetheless, the calculation of the asymmetry parameter (g-parameter),[36,37] defined as the ratio of the cosine-weighted scattering cross-section over the total scattering cross-section, shows that, in the dipolar regime, successive first Kerker conditions[38] exist for several refractive indices and a wide range of size parameters.[30,31] It was further demonstrated that the simultaneous suppression of electric and magnetic dipolar modes results in zero total scattering efficiency under illumination with a PDF across various refractive indices and a broad range of size parameters in the limit of large particles.[30] This phenomenon is associated with the excitation of non-radiating *hybrid* optical anapoles.[32,39,40] However, since illumination of particles using a PDF has not been experimentally implemented yet, alternatively, it is possible to engineer the multipolar content of the tightly-focused Gaussian beams (TFGBs)[41–45] to replicate the abovementioned scattering properties under PDF illumination.[30] Following these predictions, the dipolar regime was unveiled, and the first Kerker conditions were observed experimentally in mid-index dielectric microspheres, with diameters d ≈ 0.78 and 0.88 μm under TFGB illumination.[46]

Here, we report, for the first time, the experimental detection of hybrid optical anapoles in dielectric microspheres (Titanium dioxide, $TiO_2$) using TFGB illumination. In previous studies, detection of optical anapoles is typically performed by measuring the scattered radiation in the backward direction, with the detection angles given by the angular range corresponding to the numerical aperture (NA) of the objective lens.[12,28,29,47,48] However, detecting only the backward-propagated radiation cannot uniquely distinguish the presence of an optical anapole, as the excitation of dielectric microspheres with TFGBs can also lead to several scattering minima associated with the first Kerker condition in the backward direction.[30,31,46] To detect the presence of an anapole and confirm its non-radiating nature, scattering must be measured in additional directions, for instance in the forward direction, to ensure that the radiation completely cancels out. Typically, scattering in the forward direction is detected using dark-field scattering spectroscopy.[8,49] However, in our case, the excitation of hybrid optical anapoles in dielectric microspheres requires using a tightly-focused light beam to excite dielectric microspheres. This poses a challenge, as the inability to employ a condenser for tightly focusing the beam renders dark-field spectroscopy unsuitable for measuring scattering in the forward direction. To address this experimental limitation and facilitate the detection of the anapole in dielectric microspheres, we developed a dual-detection optical spectroscopy (DDOS) system. This system integrates an upright microscope with an inverted microscope, enabling the sequential measurement of scattering spectra in both forward and backward directions under TFGB illumination.

Our experimental and simulation results show that the excitation of $TiO_2$ microspheres (diameter, d ≈1 μm) under TFGB illumination leads to the detection of *nearly coincident* scattering minima in the forward and backward directions within specific wavelength ranges. The results of the multipolar decomposition show that these nearly coincident scattering minima in the forward and backward directions are due to vanishing electric and magnetic dipolar coefficients associated with hybrid optical anapoles, as predicted theoretically. We also explore the role of the dispersion relationships and diameter of $TiO_2$ microspheres on the positions and spacing of the scattering minima in both the forward and backward directions.

## 2. Results and Discussions

### 2.1. Theoretical Predictions

It was theoretically predicted that the excitation of dipolar regimes in a homogeneous mid-index (2 < n < 3.5) and high-index (n > 3.5) dielectric sphere under PDF illumination can excite optical anapoles for large particles[30] (i.e, with $x (= \pi d/\lambda)$-size parameter > 2). A PDF, by construction, is an electromagnetic field with only dipolar content; therefore, it can only excite dipolar modes in any kind of particles. For spherical particles, the scattering is characterized by the electric and magnetic Mie coefficients ($a_j$ and $b_j$). In particular, the dipolar moments are related to $a_1$ and $b_1$ coefficients. These coefficients fully determine the forward/backward asymmetry in the scattered field in the dipolar regime through the asymmetry parameter (g). In this respect, the ratio between the backward and forward scattering cross sections, $\sigma_b/\sigma_f$ in the dipolar regime is given by the g-parameter.[30,50]

$$\frac{\sigma_b}{\sigma_f} = \frac{1-2g}{1+2g} \quad (1)$$

In this regime, the asymmetry parameter can be expressed in terms of the Mie coefficients as,[30]

$$g = \frac{Re\left[a_1 b_1^*\right]}{|a_1|^2 + |b_1|^2} \quad (2)$$

It is possible to obtain a deeper insight of the scattering phenomena by writing the dipolar Mie coefficients in the phase notation: $a_1 = i \sin\alpha_1 e^{-i\alpha_1}$ and $b_1 = i \sin\beta_1 e^{-i\beta_1}$.[51] Here, $\alpha_1$ and $\beta_1$ are related to the Riccati-Bessel functions, as described in reference.[30] An intriguing phenomenon that emerges under dipolar field illumination is the appearance of distinct electric or magnetic scattering regimes when $\sin\beta_1 = 0$ (or $\sin\alpha_1 = 0$), which give rise to g = 0, according to Equation (2).[30] Also, this leads to the scattering efficiency approaching zero when both the electric and magnetic scattering amplitudes simultaneously vanish. Interestingly, this condition can be found for several refractive indices across a wide range of x-size parameters under dipolar excitation. We have calculated the scattering efficiency of a lossless dielectric sphere as a function of refractive index (n) and size parameter (*x*) for a plane wave illumination as shown in **Figure 1**. The calculation results show the appearance of several scattering minima associated with *g*-parameter singularities for a given multipole order satisfying, $a_1 = b_1 = 0$, $a_2 = b_2 = 0$; $a_3 = b_3 = 0$, *etc.* These *g*-parameter singularities are non-radiating spectral points, for a given multipolar mode, and are referred to as hybrid optical anapoles.[32,39,40]

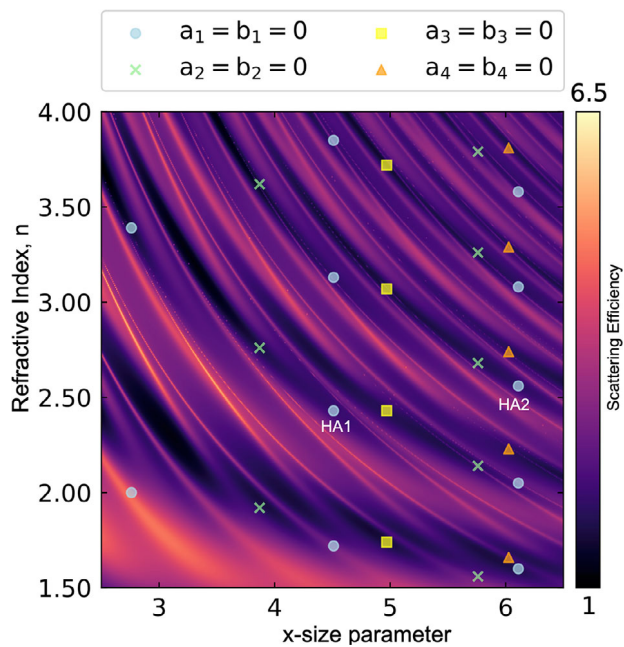

**Figure 1.** Color map showing scattering efficiency of a lossless dielectric sphere as a function of refractive index and size parameter for a plane wave illumination. The blue solid circles are associated with singularities in g parameter associated with $a_1 = b_1 = 0$, which is referred to as hybrid optical anapoles. The $a_2 = b_2 = 0$, $a_3 = b_3 = 0$, etc. are associated with higher-order hybrid optical anapoles. HA1 and HA2 are the hybrid optical anapoles experimentally detected here.

As shown in Figure 1, these hybrid optical anapoles typically emerge within a size parameter range of approximately $x \sim 2.5$ to $x \sim 6$. At larger size parameters, the excitation of multiple multipolar modes, including quadrupoles, octupoles, and higher-order terms, leads to significant spectral overlap. This makes the far-field cancellation signature of anapole states increasingly difficult to distinguish, complicating both theoretical identification and experimental observation of higher-order anapole-like modes. Moreover, material losses, dispersion, and geometry imperfections introduce additional complications that further limit the experimental observation and isolation of anapole states with larger size parameters. Note that, in theory, hybrid dipolar optical anapoles associated with the condition, $a_1 = b_1 = 0$ representing the simultaneous suppression of electric and magnetic dipole radiation, can occur for arbitrarily large size parameters, if higher-order multipolar modes (e.g., quadrupoles, octupoles) are entirely absent. In this idealized scenario, observation of anapole states at large size parameters is constrained by the asymptotic decay of the spherical Bessel functions, which render the particle effectively transparent to the incident light. Nevertheless, based on these calculation results, it should be possible to excite dipolar hybrid optical anapoles, indicated by the blue solid circles in

Figure 1, satisfying the condition $a_1 = b_1 = 0$ under PDF illumination. Note that the theoretical singularities in *g* (hybrid anapoles) plotted in Figure 1 do not yield zeros in the scattering efficiency due to the contribution of multipolar modes under plane wave illumination, for which structured light illumination is needed.

### 2.2. Dual Detection Optical Spectroscopy

From the theoretical prediction of Figure 1, it can be deduced that for a dielectric sphere of diameter, $d = 1.07$ μm with (*i*) $x = 4.52$, one can expect to observe a hybrid anapole at $\lambda = 743$ nm for $n = 2.43$ (indicated by HA1); (*ii*) $x = 6.12$, one can expect to observe a hybrid anapole at $\lambda = 549$ nm for $n = 2.56$ (indicated by HA2). Previous reports have shown that $TiO_2$ is a dispersive material with a refractive index change in the range of 2.4 to 2.6 as a function of wavelength in the optical regime.[52–56] Therefore, the refractive index range of $TiO_2$ microspheres falls within the theoretically predicted values and can be utilized to detect hybrid anapoles, HA1 and HA2, in the optical frequencies. For this purpose, we used $TiO_2$ microspheres that were synthesized using the isopropyl titanate hydrolysis method.[57]

As mentioned above, to achieve an unambiguous detection of the anapole, in addition to measuring the scattering minima in the backward direction, it is essential to measure scattering minima along additional directions. **Figure 2a** shows the schematic diagram of the DDOS setup we developed by integrating an upright microscope (Olympus IX2) on top of an inverted microscope (Olympus IX73). In this setup, the output of a white light continuum (Leukos SMHP 4.0) is coupled to an inverted optical microscope equipped with a high NA objective lens. For creating a tightly-focused beam at the sample plane, we used a 40×, NA = 0.95 objective (Olympus, UPLXAPO). The back-scattered spectra are acquired by a Charge-Coupled Device (CCD) (Andor, Newton) connected to an imaging spectrometer (Shamrock 303i) coupled to the side port of the microscope using a 4-*f* relay system. The forward scattered spectra are acquired by collecting the scattering using another 40×, NA = 0.95 objective (Olympus, UPLXAPO) in the forward direction and directing the light using a combination of mirrors, another 4-*f* relay system, and a flipper mirror into the path of the same CCD and spectrometer. The inset of Figure 2a shows an optical image of an experimentally generated tightly-focused Gaussian beam.

The diluted solution of the microspheres is drop-casted on a patterned glass substrate placed at the focal plane of both the objectives, having a common optical axis, so that the scattering in both forward and backward directions can be collected by simply switching the position of the flipper mirror (FM). The scattering spectra are measured at the single-particle level embedded in air, and the size and shape of the microspheres are determined by a correlated SEM-Optical Microscopy approach.[29,34,58] For more details on correlated single-particle spectroscopy measurements, see Section S1 (Supporting Information). Furthermore, to ensure accurate excitation of resonant modes under TFGB illumination, individual microspheres were raster-scanned across the focal plane while recording the scattering spectra. This procedure ensures that each microsphere is positioned at the focal point of the incident beam, minimizing the effects of off-center or partial illumination and enabling reliable observation of dipolar and anapole-associated features. To plot scattering spectra, each measured spectrum is normalized to a reference spectrum, a process typically carried out while performing interferometric scattering spectroscopy.[47,48] For the measurement of scattering in the backward direction, the reflected beam off the glass substrate without particles is used as a reference. Similarly, for the measurement of scattering in the forward direction, the transmitted signal without the particle is used as a reference. Note that, unlike the typical assumption for nanoparticles, in which the contributions from the interference between the reference and scattering are neglected,[47,48] for microspheres, the experimentally measured scattering spectra also contain contributions from the interference between the reference and scattering signals. However, an agreement of the positions of the scattering minima and the overall spectral shape in experimental and calculated scattering spectra confirms the dominance of the scattering signal in the measurements. More details on the contributions from the interference between the reference and scattering and normalization of the scattering spectra are discussed in Section S2 (Supporting Information).

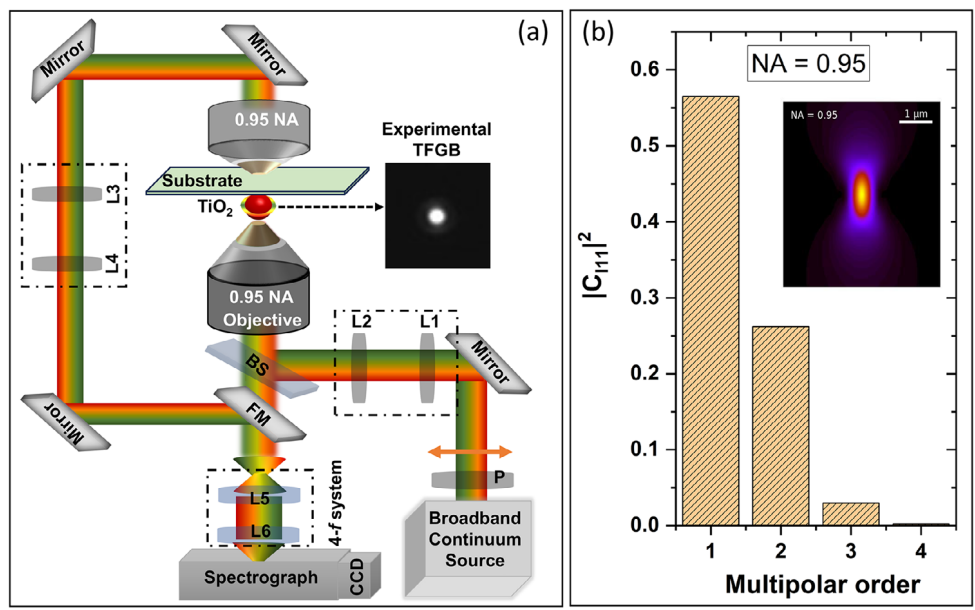

**Figure 2.** a) Schematic diagram of a dual detection optical spectroscopy (DDOS) setup developed by integrating an upright microscope on top of an inverted microscope. The back-scattered spectra are acquired by a CCD connected to an imaging spectrometer coupled to the side port of the microscope. The forward-scattered spectra are acquired by collecting the scattering using another objective in the forward direction and directing the light into the path of the same CCD and spectrometer. The inset of Figure 2a shows an optical image of an experimentally generated tightly-focused Gaussian beam. (L → Lens, BS → Beam Splitter, FM → Flipper Mirror). b) Calculated weights of each $C_{l11}$ multipolar order, and the inset shows an image of a numerically calculated tightly-focused Gaussian incident beam with NA = 0.95, propagating along the z-direction.

As mentioned above, it has been demonstrated that TFGBs can attain a scattering situation close to dipolar field illumination[30] and, thereby, can be used to select a few relevant Mie coefficients and control the relative weight of the different multipolar modes (given by $a_l$ and $b_l$) with the use of the multipolar content of the incident field with beam shape coefficients (BSC) $C_{lm_z\sigma}$.[45] The BSCs are the coefficients of the multipolar modes of order $l$, with total angular momentum in the z-direction $m_z$ and helicity $\sigma$. Figure 2b shows the calculated weights of the relevant $C_{l11}$ for a TFGB focused under our experimental conditions (NA = 0.95).

The inset shows an image of the propagating tightly-focused Gaussian incident beam in the vertical (z)direction. The value of the beam shape coefficient, $C_{l11}$, heavily decreases as the multipolar order $l$ increases. In fact, for NA = 0.95, the relative weight of the dipolar content of the incident beam is ≈80%, and the quadrupolar contribution and higher multipolar orders represent ≈20% (which may vary depending on the diameter of the beam) of the content of the incident beam.[46] Therefore, the TFGBs are dominantly dipolar in nature, and they can be used to excite hybrid optical anapoles in $TiO_2$ microspheres. Note that the supercontinuum laser exhibits different laser spot sizes as a function of wavelength. We have taken this into account while calculating the beam shape coefficients as described elsewhere.[59] Within our measurement window (480 – 800 nm), the wavelength-dependent spot size was found to be negligible.

### 2.3. Forward and Backward Scattering Spectra of the Microspheres

The experimental forward and backward scattering spectra measured using DDOS spectroscopy are shown in **Figure 3**a,b for single $TiO_2$ microspheres with $d \approx 1.07\pm0.03$, and $d \approx 1.02\pm0.02$ μm, respectively. The diameters of the single microspheres were determined from the correlated SEM-Optical microscopy, with the insets showing the SEM image of the exact same microspheres whose scattering spectra were measured. Since the microspheres are not perfectly spherical, the average diameters were determined by measuring the diameters in different directions. The experimental results are complemented by numerical simulations, as shown in Figure 3c,d for $TiO_2$ microspheres with $d = 1.07$, and $d = 1.02$ μm, respectively. We have included color maps showing simulated forward and backward scattering intensity of the TiO$_2$ microspheres as a function of diameter in Section S3 (Supporting Information). The calculations were performed using the generalized Lorenz–Mie theory (GLMT).[45] The electric and magnetic Mie coefficients were taken from Bohren and Huffmann[60] to solve the exact scattering problem of a tightly-focused Gaussian beam impinging on a homogeneous dielectric sphere. Then, the resulting electromagnetic field intensity was integrated for the angular range corresponding to the NA of the objective used in the experiment (NA = 0.95) to calculate the forward and backward scattering spectra.

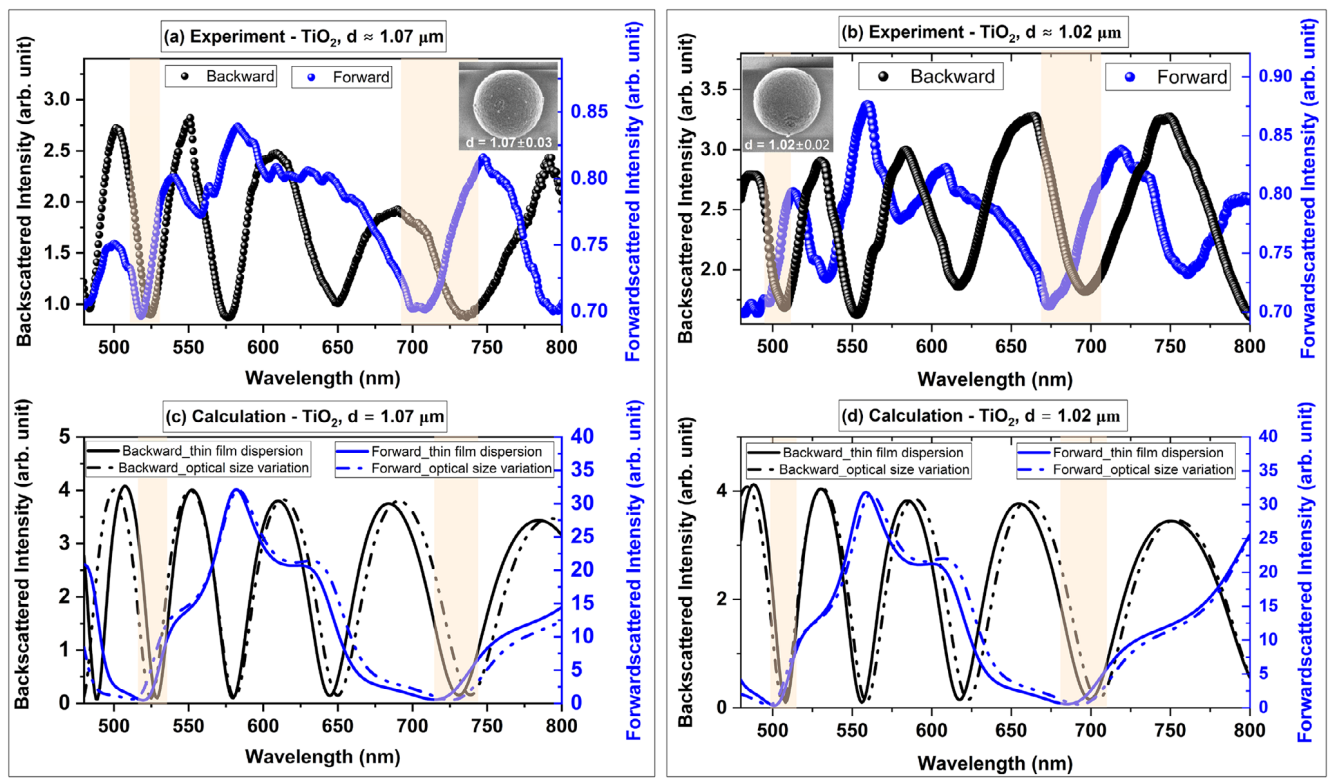

**Figure 3.** Experimental (a,b) and calculated (c,d) scattering spectra of a single TiO$_2$ microsphere with $d \approx 1.07$ μm (a, c), and $d \approx 1.02$ μm (b, d) in the forward and backward directions under TFGB illumination. The insets of (a) and (b) show the SEM image of the exact same microspheres whose scattering spectra were measured. Both the experimental and simulated scattering spectra show the appearance of several scattering minima associated with the first Kerker condition in the backward direction. In the forward direction, two distinct scattering minima appear in both the experimental and calculated scattering spectra, which *nearly* coincide with two scattering minima in the backward direction, as indicated by the light orange highlighted regions. These coincidences in the forward and backward scattering minima are due to the excitation of hybrid optical anapoles in the TiO$_2$ microspheres.

Both the experimental and simulated scattering spectra show the appearance of several scattering minima in the backward direction for TiO$_2$ microspheres with $d \approx 1.07$, and $d \approx 1.02$ μm. These scattering minima are associated with the first Kerker condition when $a_1 = b_1$ is satisfied, as it has been theoretically predicted.[30] On the other hand, in the forward direction, two distinct scattering minima appear in both the experimental and simulated scattering spectra within the measured and calculated wavelength range. These two scattering minima in the forward direction are found to *nearly* coincide with two scattering minima in the backward direction, as indicated by the orange-highlighted region in both the experimental and simulated spectra in Figure 3. These coincidences of the scattering minima in the forward and backward scattering spectra can be attributed to the excitation of hybrid optical anapoles, HA1 and HA2, as predicted theoretically (Figure 1). This is further supported by the multipolar decomposition of the scattering spectra and the calculation of Mie coefficients, as we will later discuss in Section 2.5 below.

Even though the positions and spacings of the scattering minima and the general trends of the experimentally measured and simulated spectra agree well, we would like to point out that the dispersion of the TiO$_2$ microspheres plays a critical role in determining the overall shape of the scattering spectra. A detailed discussion on the role of dispersion on the optical spectra is presented in Section S4 (Supporting Information). Briefly, the positions of the scattering minima, the spacings amongst them, the scattering intensity, and thereby, the overall shape of the scattering spectra, heavily depend on the dispersion relationships of TiO$_2$ among various morphological phases, such as anatase nanoparticles,[52] and thin films.[53,54,56] In our simulations, the spectra calculated using a thin film dispersion relationship determined by Zhukovsky and co-workers[56] best agreed with the experimental results and are plotted as solid curves in Figure 3c,d. To further validate the comparison of the experimental and simulation results, we determined the variation of optical size or the refractive index contrast by fitting the backscattered spectra of the microspheres in terms of a sinusoidal function with regular oscillations dependent on its optical size,[61] and calculated the forward and backscattered spectra as shown by the dashed dot curves in Figure 3c,d. Nevertheless, calculated spectra using both dispersion relations agree well with the experimentally measured spectra. The larger deviation in forward scattering is primarily due to the stronger incident field in the forward direction, which enhances the interference effects. The detected signal resulting from the superposition of incident, transmitted, and scattered waves produces interference effects that are highly sensitive to optical alignment, substrate reflections, and the numerical aperture of the collection optics.

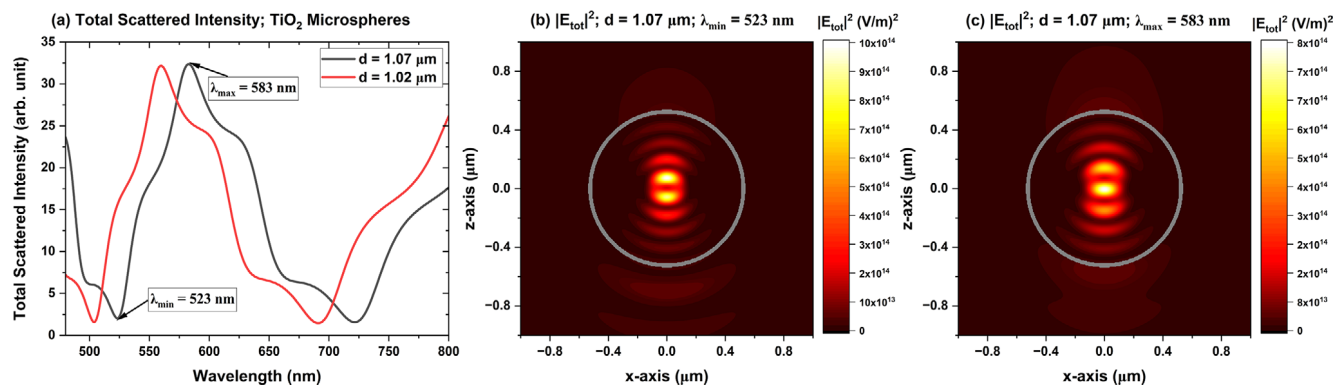

**Figure 4.** a) shows the calculated total scattering spectra for the TiO$_2$ microspheres with $d = 1.07$ μm, and $d = 1.02$ μm. b) and c) show the spatial total electric field distribution, in the x-z plane (at y = 0) of the TiO$_2$ microspheres with $d = 1.07$ μm, at two wavelengths: $\lambda_{min} \approx 523$ and $\lambda_{max} \approx 583$ nm, respectively. At 523 nm (b), the internal field exhibits a highly confined, symmetric distribution with pronounced internal nodes, characteristic of a high-order standing wave and consistent with interference patterns between the internal field components.

### 2.4. Total Scattering Spectra and Internal Field Distributions

The calculated total scattering spectra, obtained by summing the forward and backward scattering intensities for the TiO$_2$ microspheres with $d = 1.07$, and $1.02$ μm, are shown in **Figure 4**a. For $d = 1.07$ μm two scattering minima associated with the excitation of hybrid anapoles appear at $\lambda_{min} \approx 523$ and 721 nm, with the maximum total scattering intensity occurring at $\lambda_{max} \approx 583$ nm. The spatial total electric field distributions of the TiO$_2$ microspheres with $d = 1.07$ μm, at two wavelengths: $\lambda_{min} \approx 523$ and $\lambda_{max} \approx 583$ nm, as shown in Figure 4b,c, respectively, show subtle differences in spatial localization. At 523 nm (Figure 4b), the internal field exhibits a highly confined, symmetric distribution with pronounced internal nodes, characteristic of a high-order standing wave and consistent with interference patterns between the internal field components. In contrast, the field at 583 nm (Figure 4c) is more diffuse, with weaker spatial confinement. These standing wave features are governed by the spherical Bessel function of the first kind and first order, $j_1(nkr)$ (where $k$ is the wave number and $r$ is the radial distance from the center of the particle), which defines the radial profile of the dipolar mode. The *radial order* of a mode is defined by the number of internal *nodes* (zero crossings of $j_1(nkr)$). In our case, the internal radial field distribution follows $|E_r(r)| \propto |j_1(nkr)|$. For a TiO$_2$ sphere of diameter $d = 1.07$ μm, the field exhibits four nodes at $\lambda_{min} = 523$ nm, as shown in Figure 4b. Also, while the field magnitudes outside the particle are comparable at both wavelengths, the spatial distribution of the external field at 583 nm is broader, suggesting a more radiative character compared to the more localized near-field observed at 523 nm.

These results indicate that the hybrid anapole exhibits enhanced internal field confinement and a more spatially localized external field. However, the near-field patterns and magnitudes at the scattering maximum and minimum are similar, suggesting that any differences are subtle and not readily apparent from field plots alone. Thus, further analysis, such as calculation of Mie coefficients, is necessary to determine whether it corresponds to an anapole or other radiating state, as detailed in the following sections.

Note that, for the excitation of an *ideal* hybrid optical anapole, one would expect the total scattering intensity (Figure 4a) to completely vanish, as well as the scattering minima in both the forward and backward directions (Figure 3) to *completely* coincide with each other and vanish. As discussed subsequently in sections, the contributions from higher-order multipole and the in-exact value of the refractive index of the TiO$_2$ microsphere are responsible for this deviation from the theoretical prediction.

### 2.5. Multipolar Decomposition of the Scattering Spectra and Mie Coefficients

To verify that these coincidences in forward and backward scattering minima shown in Figure 3 are indeed due to the excitation of hybrid optical anapoles in the TiO$_2$ microspheres, we have performed multipolar decomposition of the backward and forward scattering spectra, as well as calculated the magnitude of the dipolar electric and magnetic Mie coefficients. **Figure 5** shows the multipolar decomposition of the forward and backward scattering spectra of Figure 3c, for TiO$_2$ microspheres with $d = 1.07$ μm. The multipolar decomposition of the scattering spectra of Figure 3d, for TiO$_2$ microspheres with $d = 1.02$ μm is shown in Section S5 (Supporting Information). The results show that the scattering spectra in both the forward and backward directions are dominated by the dipolar contributions (both electric and magnetic). However, there is a significant contribution from the quadrupolar response, especially in the backscattered spectra, and there is some contribution from the octupolar response due to the multipolar content associated with the TFGB excitation (Figure 2b). These multipolar contributions shift the wavelengths as well as offset the intensity of the scattering minima that appear due to dipolar excitation, which prevents the total scattering intensity ($Q_{tot}$) from completely vanishing at the first Kerker conditions and anapole excitation.

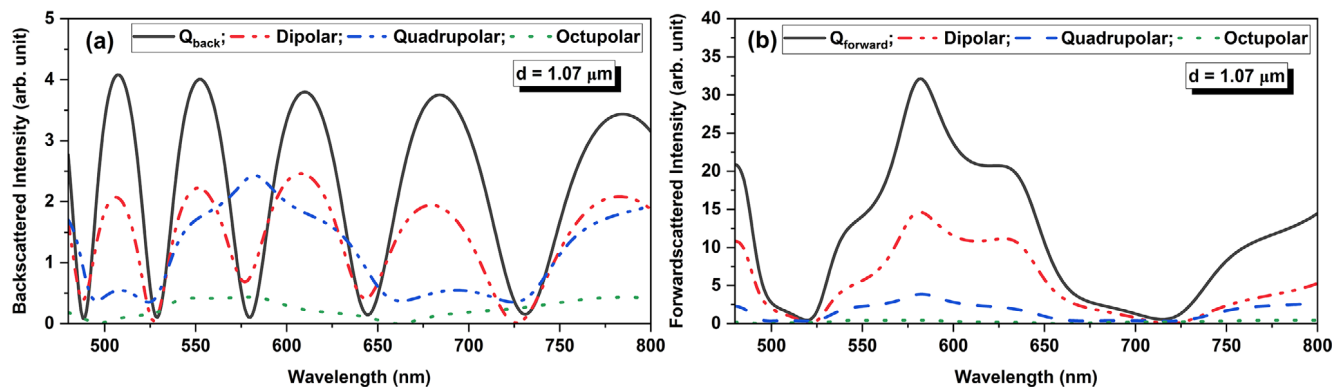

**Figure 5.** Multipolar decomposition of the scattering spectra in the backward a) and forward b) directions for a TiO$_2$ microsphere with $d \approx 1.07$ μm under TFGB illumination. The scattering spectra in both the forward and backward directions are dominated by the dipolar contributions (electric and magnetic), with non-negligible contributions from higher-order modes.

The magnitudes of the sum of the dipolar electric and magnetic Mie coefficients, $|a_1 + b_1|^2$, and their difference, $|a_1 - b_1|^2$ are plotted in **Figure 6** for TiO$_2$ spheres with the dispersion relations used to calculate the solid curve in Figure 3c. In the forward direction ($\theta = 0°$), the scattering intensity is proportional to $|a_1 + b_1|^2$, whereas the backward direction ($\theta = 180°$), the scattering intensity is proportional to $|a_1 - b_1|^2$.[60] Therefore, $|a_1 + b_1|^2 = |a_1 - b_1|^2 = 0$ can only be possible when $a_1 = b_1 = 0$, which, as mentioned earlier, is the condition for hybrid optical anapoles. As shown in **Figure 6**, $|a_1 + b_1|^2 = |a_1 - b_1|^2 = 0$ appear within the highlighted regions, where the electric and magnetic dipolar coefficients nearly coincide with each other. Hence, we can conclude that these coincidences in the forward and backward scattering minima of Figure 3 are due to the excitation of hybrid optical anapoles in the TiO$_2$ microspheres.

### 2.6. Factors Contributing to Non-Ideal Excitation of Optical Anapoles

As mentioned earlier, for the excitation of an ideal hybrid anapole, one would expect the intensity of the scattering minima in both the forward and backward directions to completely vanish and coincide with each other. However, two major factors are responsible for this *non-ideal* excitation of the hybrid anapoles – (*i*) Contributions from higher order modes, such as quadrupolar and octupolar, appear within the scattering spectra, as shown in Figure 5. These multipolar contributions shift the wavelengths as well as offset the intensity of the scattering minima, which prevents the total scattering intensity from completely vanishing at the anapole wavelength. (*ii*) Small differences in the index of refraction due to, for example, dispersion. The appearance of ideal optical anapoles is associated to specific refractive indices and size parameters, as shown in Figure 1. However, the TiO$_2$ microspheres studied here can possess a dispersion relationship with the refractive index in the range of 2.4 to 2.6 as a function of wavelength in the optical regime.[52–56] The position and spacing of the scattering minima heavily depend on the dispersion relations, as discussed in Section 2.3 above and Section S4 (Supporting Information). The TiO$_2$ microspheres studied here do not exactly satisfy the exact relationship between the refractive index and size parameter (as indicated in Figure 1), slightly displacing the scattering minima in both the forward and backward directions and preventing them from completely coinciding with each other.

To verify that these two factors are indeed responsible for the *non-ideal* excitation of the hybrid anapoles, we have calculated the *dipolar* (no higher order contributions) forward and backward scattering spectra for TiO$_2$ spheres with $d = 1.07$ μm, and exact (non-dispersive) refractive index, $n = 2.43$ and 2.56, as shown in **Figure 7a,b**, respectively. The results show that the forward and backward scattering intensities, and $|a_1 + b_1|^2$ and $|a_1 - b_1|^2$ completely vanish and coincide at $\lambda = 743$ nm for $n = 2.43$, and at $\lambda = 549$ nm for $n = 2.56$, leading to the excitation of ideal hybrid optical anapoles, as predicted theoretically and indicated by HA1 and HA2, respectively in Figure 1. The magnitudes of the sum of the dipolar electric and magnetic Mie coefficients, $|a_1 + b_1|^2$, and their difference, $|a_1 - b_1|^2$ for TiO$_2$ spheres with $d = 1.07$ μm, and $n = 2.43$ and 2.56 are presented in Section S6 (Supporting Information).

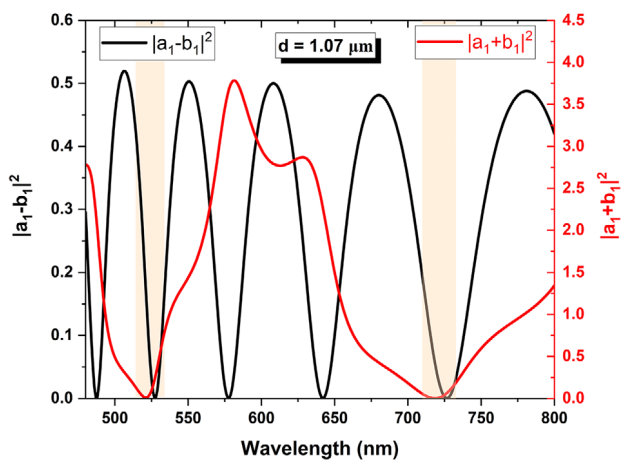

**Figure 6.** The Sum and difference of the dipolar electric ($a_1$) and magnetic ($b_1$) Mie coefficients. The scattering minima associated with $|a_1 + b_1|^2 = |a_1 - b_1|^2 = 0$ can only be possible when $a_1 = b_1 = 0$, which is referred to as hybrid optical anapoles.

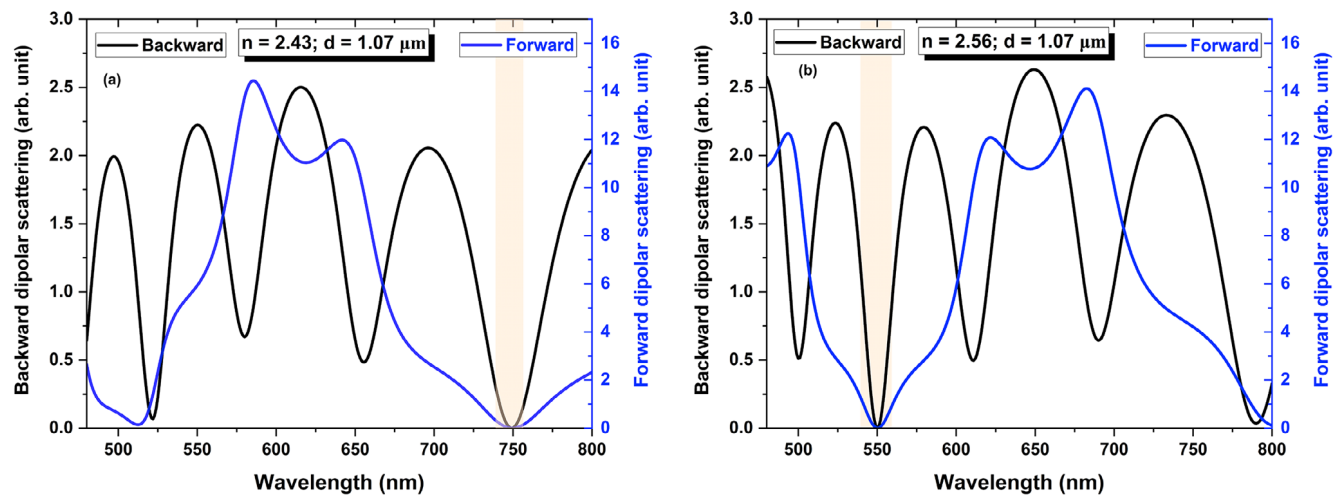

**Figure 7.** The dipolar forward and backward scattering spectra for *non-dispersive* TiO$_2$ spheres with $d = 1.07$ μm, and $n = 2.43$ (a) and 2.56 (b), respectively.

The surrounding refractive index plays a critical role in determining the anapole resonance by modifying the effective size parameter and shifting the argument of the spherical Bessel function. These changes affect both the resonance condition and the spatial field distribution inside the particle. In practical implementations, the presence of a substrate contributes to this surrounding refractive environment and can further introduce symmetry-breaking and index contrast effects that perturb the modal structure and scattering behavior of dielectric particles.[62–65] However, in our experimental configuration, the resonant modes are predominantly confined within the particle (as shown in Figure 4b,c), and interference between incident, reflected, and transmitted waves makes it difficult to isolate substrate-specific contributions. However, given the good agreement between theoretical and experimental forward and backward scattering spectra (Figure 3), the impact of substrate-induced effects is expected to be minimal.

While the ideal anapole condition requires precise tuning of the microsphere size and index, our results show that even in non-ideal scenarios, significant scattering suppression can be achieved. For example, previous calculations showed[46] that for TiO$_2$ microspheres, the internal energy density can be enhanced by a factor of ≈7.5 despite slight deviations from the optimal size and refractive index. For practical applications, achieving uniform microsphere size and refractive index is essential to realize the anapole resonance wavelength. Homogeneous dielectric microspheres can be fabricated using sol–gel methods[66] or flame aerosol synthesis,[67] with post-processing techniques such as centrifugation, improving size uniformity.[68] Refractive index tuning can also be achieved through block copolymer self-assembly[69] or electrical modulation techniques,[70] enabling scalable and application-ready implementations.

## 3. Conclusion

In conclusion, we have experimentally detected hybrid optical anapoles in dielectric microspheres (Titanium dioxide, TiO$_2$) using dual detection optical spectroscopy under tightly focused Gaussian beam (TFGB) illumination for the first time. The dual detection spectroscopy was performed by integrating an upright microscope on top of an inverted microscope and measuring the scattering spectra of TiO$_2$ dielectric microspheres in both the forward and backward directions. Several scattering minima associated with the first Kerker condition appear in the backward scattering spectra. On the other hand, in the forward direction, two distinct scattering minima appear in the scattering spectra, which nearly coincide with two scattering minima in the backward direction. These coincidences of the scattering minima in the forward and backward scattering spectra can be attributed to the excitation of hybrid optical anapoles in the dielectric microspheres. However, factors such as contributions from higher-order multipolar modes and the dispersive nature of the dielectric materials heavily influence the positions, spacing, and intensity of the scattering minima, resulting in a non-ideal excitation of the hybrid anapoles. The dielectric microspheres possessing a non-radiating anapole state in the mid-index regime can be potentially used as building blocks to form metasurfaces with high-quality factors and overcome non-efficient coupling and non-directionality of light.

## Supporting Information

Supporting Information is available from the Wiley Online Library or from the author.

## Acknowledgements

U.M. and I.G.-V. contributed equally to this work. This material is based upon work supported by the National Science Foundation (NSF), Division of Materials Research under Grant No. DMR-2208240 and DMR-2116612. G.M.-T., I.G.-V., and I.T. acknowledge support from CSIC Research Platform on Quantum Technologies PTI-001, from IKUR Strategy under the collaboration agreement between Ikerbasque Foundation and DIPC/MPC on behalf of the Department of Education of the Basque Government and from Project No. PID2022-143268NB-I00 of Ministerio de Ciencia, Innovación y Universidades.

## Conflict of Interest

The authors declare no conflict of interest.

## Data Availability Statement

The data that support the findings of this study are available from the corresponding author upon reasonable request.

## Keywords

hybrid anapole, microphotonics, mie scattering, nanophotonics, optical anapole, optical spectroscopy

Received: April 25, 2025
Revised: June 3, 2025
Published online: August 5, 2025

[1] K. Kim, E. Wolf, *Opt. Commun.* **1986**, *59*, 1.
[2] A. J. Devaney, E. Wolf, *Phys. Rev. D* **1973**, *8*, 1044.
[3] K. V. Baryshnikova, D. A. Smirnova, B. S. Luk'yanchuk, Y. S. Kivshar, *Adv. Opt. Mater.* **2019**, *7*, 1801350.
[4] V. Savinov, N. Papasimakis, D. P. Tsai, N. I. O. Zheludev, *Commun. Phys.* **2019**, *2*, 69.
[5] K. Koshelev, G. Favraud, A. Bogdanov, Y. Kivshar, A. Fratalocchi, *Nanophotonics* **2019**, *8*, 725.
[6] Y. Yang, S. Bozhevolnyi, *Nanotechnology* **2019**, *30*, 204001.
[7] A. Valero, E. Gurvitz, F. Benimetskiy, D. Pidgayko, A. Samusev, A. Evlyukhin, V. Bobrovs, D. Redka, M. Tribelsky, M. Rahmani, K. Kamali, A. Pavlov, A. Miroshnichenko, A. Shalin, *Laser Photonics Rev.* **2021**, *15*, 2100114.
[8] A. E. Miroshnichenko, A. B. Evlyukhin, Y. F. Yu, R. M. Bakker, A. Chipouline, A. I. Kuznetsov, B. Luk'yanchuk, B. N. Chichkov, Y. S. Kivshar, *Nat. Commun.* **2015**, *6*, 8069.
[9] V. A. Fedotov, A. V. Rogacheva, V. Savinov, D. P. Tsai, N. I. Zheludev, *Sci. Rep.* **2013**, *3*, 2967.
[10] M. Kerker, *J. Opt. Soc. Am.* **1975**, *65*, 376.
[11] W.-C. Zhai, T.-Z. Qiao, D.-J. Cai, W.-J. Wang, J.-D. Chen, Z.-H. Chen, S.-D. Liu, *Opt. Express* **2016**, *24*, 27858.
[12] G. Grinblat, Y. Li, M. P. Nielsen, R. F. Oulton, S. A. Maier, *ACS Nano* **2017**, *11*, 953.
[13] G. Grinblat, Y. Li, M. P. Nielsen, R. F. Oulton, S. A. Maier, *Nano Lett.* **2016**, *16*, 4635.
[14] L. Xu, M. Rahmani, K. Z. Kamali, A. Lamprianidis, L. Ghirardini, J. Sautter, R. Camacho-Morales, H. T. Chen, M. Parry, I. Staude, G. Q. Zhang, D. Neshev, A. E. Miroshnichenko, *Light-Sci. Appl.* **2018**, *7*, 44.
[15] M. R. Shcherbakov, D. N. Neshev, B. Hopkins, A. S. Shorokhov, I. Staude, E. V. Melik-Gaykazyan, M. Decker, A. A. Ezhov, A. E. Miroshnichenko, I. Brener, A. A. Fedyanin, Y. S. Kivshar, *Nano Lett.* **2014**, *14*, 6488.
[16] J. S. T. Gongora, A. E. Miroshnichenko, Y. S. Kivshar, A. Fratalocchi, *Nat. Commun.* **2017**, *8*, 15535.
[17] T. H. Feng, Y. Xu, W. Zhang, A. E. Miroshnichenko, *Phys. Rev. Lett.* **2017**, *118*, 173901.
[18] R. Wang, L. Dal Negro, *Opt. Express* **2016**, *24*, 19048.
[19] M. Gupta, Y. K. Srivastava, M. Manjappa, R. Singh, *Appl. Phys. Lett.* **2017**, *110*, 121108.
[20] A. K. Ospanova, I. V. Stenishchev, A. A. Basharin, *Laser Photonics Rev.* **2018**, *12*, 1800005.
[21] S. D. Liu, Z. X. Wang, W. J. Wang, J. D. Chen, Z. H. Chen, *Opt. Express* **2017**, *25*, 22375.
[22] A. A. Basharin, V. Chuguevsky, N. Volsky, M. Kafesaki, E. N Economou, *Phys. Rev. B* **2017**, *95*, 035104.
[23] Y. Q. Yang, V. A. Zenin, S. I. Bozhevolnyi, *ACS Photonics* **2018**, *5*, 1960.
[24] Y. B. Zhang, W. W. Liu, Z. C. Li, Z. Li, H. Cheng, S. Q. Chen, J. G. Tian, *Opt. Lett.* **2018**, *43*, 1842.
[25] G. Grinblat, Y. Li, M. P. Nielsen, R. F. Oulton, S. A. Maier, *ACS Photonics* **2017**, *4*, 2144.
[26] D. Baranov, R. Verre, P. Karpinski, M. Käll, *ACS Photonics* **2018**, *5*, 2730.
[27] T. Green, D. Baranov, B. Munkhbat, R. Verre, T. Shegai, M. Käll, *Optica* **2020**, *7*, 680.
[28] T. Zhang, Y. Che, K. Chen, J. Xu, Y. Xu, T. Wen, G. Lu, X. Liu, B. Wang, X. Xu, Y. Duh, Y. Tang, J. Han, Y. Cao, B. Guan, S. Chu, X. Li, *Nat. Commun.* **2020**, *11*, 3027.
[29] J. A. Parker, H. Sugimoto, B. Coe, D. Eggena, M. Fujii, N. F. Scherer, S. K. Gray, U. Manna, *Phys. Rev. Lett.* **2020**, *124*, 097402.
[30] C. Sanz-Fernandez, M. Molezuelas-Ferreras, J. Lasa-Alonso, N. de Sousa, X. Zambrana-Puyalto, J. Olmos-Trigo, *Laser Photonics Rev.* **2021**, *15*, 2100035.
[31] J. Olmos-Trigo, D. R. Abujetas, C. Sanz-Fernandez, X. Zambrana-Puyalto, N. de Sousa, J. A. Sanchez-Gil, J. J. Saenz, *Phys. Rev. Res.* **2020**, *2*, 043021.
[32] B. Luk'yanchuk, R. Paniagua-Dominguez, A. I. Kuznetsov, A. E. Miroshnichenko, Y. S Kivshar, *Phys. Rev. A* **2017**, *95*, 063820.
[33] P. Wozniak, P. Banzer, G. Leuchs, *Laser Photonics Rev.* **2015**, *9*, 231.
[34] U. Manna, H. Sugimoto, D. Eggena, B. Coe, R. Wang, M. Biswas, M. Fujii, *J. Appl. Phys.* **2020**, *127*, 033101.
[35] X. Zambrana-Puyalto, X. Vidal, P. Wozniak, P. Banzer, G. Molina-Terriza, *ACS Photonics* **2018**, *5*, 2936.
[36] J. Olmos-Trigo, *ACS Photonics* **2024**, *11*, 3697.
[37] J. Olmos-Trigo, *Nano Lett.* **2024**, *24*, 8658.
[38] M. Kerker, D.-S. Wang, C. L. Giles, *J. Opt. Soc. Am.* **1983**, *73*, 765.
[39] P. Kapitanova, E. Zanganeh, N. Pavlov, M. Z. Song, P. Belov, A. Evlyukhin, A. Miroshnichenko, *Ann. Phys.* **2020**, *532*, 2000293.
[40] E. A. Gurvitz, K. S. Ladutenko, P. A. Dergachev, A. B. Evlyukhin, A. E. Miroshnichenko, A. S. Shalin, *Laser Photonics Rev.* **2019**, *13*, 1800266.
[41] L. Allen, M. W. Beijersbergen, R. J. C. Spreeuw, J. P Woerdman, *Phys. Rev. A* **1992**, *45*, 8185.
[42] F. Gori, *Opt. Commun.* **1994**, *107*, 335.
[43] N. M. Mojarad, V. Sandoghdar, M. Agio, *J. Opt. Soc. Am. B* **2008**, *25*, 651.
[44] S. Orlov, U. Peschel, T. Bauer, P. Banzer, *Phys. Rev. A* **2012**, *85*, 063825.
[45] X. Zambrana-Puyalto, X. Vidal, G. Molina-Terriza, *Opt. Express* **2012**, *20*, 24536.
[46] B. Coe, J. Olmos-Trigo, D. Qualls, M. Alexis, M. Szczerba, D. Abujetas, M. Biswas, U. Manna, *Adv. Opt. Mater.* **2023**, *11*, 2202140.
[47] K. Lindfors, T. Kalkbrenner, P. Stoller, V. Sandoghdar, *Phys. Rev. Lett.* **2004**, *93*, 037401.
[48] M. van Dijk, A. Tchebotareva, M. Orrit, M. Lippitz, S. Berciaud, D. Lasne, B. Cognet, B. Lounis, *Phys. Chem. Chem. Phys.* **2006**, *8*, 3486.
[49] V. Zenin, A. Evlyukhin, S. Novikov, Y. Yang, R. Malureanu, A. Lavrinenko, B. Chichkov, S. Bozhevolnyi, *Nano Lett.* **2017**, *17*, 7152.
[50] J. Olmos-Trigo, C. Sanz-Fernandez, F. S. Bergeret, J. J. Saenz, *Opt. Lett.* **2019**, *44*, 1762.
[51] H. C. Hulst, H. C. van de Hulst, *Light Scattering by Small Particles*, Courier Corporation, North Chelmsford, Massachusetts **1957**.

[52] I. Bodurov, I. Vlaeva, A. Viraneva, T. Yovcheva, S. Sainov, *Nanosci. Nanotechnol* **2016**, *16*, 31.

[53] A. Jolivet, C. Labbé, C. Frilay, O. Debieu, P. Marie, B. Horcholle, F. Lemarié, X. Portier, C. Grygiel, S. Duprey, W. Jadwisienczak, D. Ingram, M. Upadhyay, A. David, A. Fouchet, U. Lüders, J. Cardin, *Appl. Surf. Sci.* **2023**, *608*, 155214.

[54] J. Kischkat, S. Peters, B. Gruska, M. Semtsiv, M. Chashnikova, M. Klinkmüller, O. Fedosenko, S. Machulik, A. Aleksandrova, G. Monastyrskyi, Y. Flores, W. Masselink, *Appl. Opt.* **2012**, *51*, 6789.

[55] M. N. Polyanskiy, *Sci. Data* **2024**, *11*, 94.

[56] S. V. Zhukovsky, A. Andryieuski, O. Takayama, E. Shkondin, R. Malureanu, F. Jensen, A. V. Lavrinenko, *Phys. Rev. Lett.* **2015**, *115*, 177402.

[57] https://www.epruibiotech.com/product/tio2-microspheres/ (accessed July 2025).

[58] U. Manna, J. H. Lee, T. S. Deng, J. Parker, N. Shepherd, Y. Weizmann, N. F. Scherer, *Nano Lett.* **2017**, *17*, 7196.

[59] X. Zambrana-Puyalto, arXiv:1502.01648 **2016**, http://arxiv.org/abs/1502.01648.

[60] *Absorption and Scattering of Light by Small Particles*, Wiley, New York, Chichester, UK **1983**.

[61] M. Molezuelas-Ferreras, Á. Nodar, M. Barra-Burillo, J. Olmos-Trigo, J. Lasa-Alonso, I. Gómez-Viloria, E. Posada, J. J. M. Varga, R. Esteban, J. Aizpurua, L. E. Hueso, C. Lopez, G. Molina-Terriza, *Laser Photonics Rev.* **2024**, *18*, 2300665.

[62] A. E. Miroshnichenko, A. B. Evlyukhin, Y. S. Kivshar, B. N. Chichkov, *ACS Photonics* **2015**, *2*, 1423.

[63] D. W. Mackowski, *J. Quant. Spectrosc. Radiat. Transfer* **2008**, *109*, 770.

[64] A. B. Vasista, E. J. C. Dias, F. J. G. de Abajo, W. L. Barnes, *Nano Lett.* **2022**, *22*, 6737.

[65] J. van de Groep, A. Polman, *Opt. Express* **2013**, *21*, 26285.

[66] C. J. Brinker, G. W. Scherer, *Sol-Gel Science: The Physics and Chemistry of Sol–Gel Processing*, Academic Press, Cambridge, Massachusetts **1990**.

[67] R. Strobel, S. Pratsinis, *J. Mater. Chem.* **2007**, *17*, 4743.

[68] Y. Xia, B. Gates, Y. Yin, Y. Lu, *Adv. Mater.* **2000**, *12*, 693.

[69] J. Kim, H. Kim, B. Kim, T. Chang, J. Lim, H. Jin, J. Mun, Y. Choi, K. Chung, J. Shin, S. Fan, S. Kim, *Nat. Commun.* **2016**, *7*, 12911.

[70] X. Gao, L. Xie, J. Zhou, *Sci. Rep.* **2022**, *12*, 10117.